\documentclass[prd, superscriptaddress, tightenlines, longbibliography, nofootinbib, eqsecnum, amsfonts, amsmath, floatfix, twocolumn, notitlepage]{revtex4-2}
\pdfoutput=1
\usepackage{hyperref}
\usepackage{amsmath}
\usepackage{graphicx}
\usepackage{bm}
\usepackage[usenames,dvipsnames,svgnames,table]{xcolor}
\definecolor{linkcolor}{rgb}{0.6,0,0}
\definecolor{citecolor}{rgb}{0,0,0.75}
\definecolor{urlcolor}{rgb}{0.12,0.46,0.7}

\usepackage[capitalize]{cleveref}
\Crefname{equation}{Equation}{Equations}
\crefname{equation}{Eq.}{Eqs.}
\Crefname{figure}{Figure}{Figures}
\crefname{figure}{Fig.}{Figs.}
\Crefname{table}{Table}{Tables}
\crefname{table}{Tab.}{Tabs.}
\Crefname{section}{Section}{Sections}
\crefname{section}{Sec.}{Secs.}
\Crefname{subsection}{Subsection}{Subsections}

\newcommand{\planck}{{\it Planck }}
\newcommand{\Alens}{{A_{\rm lens}}}
\newcommand{\Sussex}{Department of Physics \& Astronomy, University of Sussex, Brighton BN1 9QH, UK}

\begin{document}

\title{CMB constraints on the early universe independent of late time cosmology}

\author{Pablo Lemos}
\affiliation{\Sussex}

\author{Antony Lewis}
\affiliation{\Sussex}

\begin{abstract}
The CMB is a powerful probe of early-universe physics but is only observed after passing through large-scale structure, which changes the observed spectra in important model-dependent ways.
This is of particular concern given recent claims of significant discrepancies with low redshift data sets when a standard $\Lambda$CDM model is assumed.
By using empirical measurements of the CMB lensing reconstruction, combined with weak priors on the smoothness of the lensing spectrum, foregrounds, and shape of any additional integrated Sachs-Wolfe effect, we show how the early-universe parameters can be constrained from CMB observations almost independently of the late-time evolution.
This provides a way to test new models for early-universe physics, and measure early-universe parameters, independently of late-time cosmology. Using the empirical measurement of lensing keeps the size of the effect of late-time modelling uncertainty under control, leading to only modest increases in error bars of most early-universe parameters compared to assuming a full evolution model. We provide robust constraints on early-$\Lambda$CDM model parameters using the latest Planck PR4 data and show that with future data marginalizing over a single lensing amplitude parameter is sufficient to remove sensitivity to late-time cosmological model only if the spectral shape matches predictions.
\end{abstract}
\maketitle

%\begin{keywords}
%keyword1 -- keyword2 -- keyword3
%\end{keywords}

%%%%%%%%%%%%%%%%%%%%%%%%%%%%%%%%%%%%%%%%%%%%%%%%%%

%%%%%%%%%%%%%%%%% BODY OF PAPER %%%%%%%%%%%%%%%%%%

\section{Introduction}

Observations of the Cosmic Microwave Background (CMB) have served to establish the flat Lambda-Cold Dark Matter ($\Lambda {\rm CDM}$) cosmological model, as the best fit to existing observations. In this model, the Universe starts with an inflationary phase, which creates the seeds for structure formation, and at present times it is dominated by a cosmological constant ($\Lambda$) driving the observed accelerated expansion, and Cold Dark Matter (CDM) as the predominant matter component. The most accurate parameter constraints within the $\Lambda {\rm CDM}$ cosmology come from CMB observations from the {\it Planck} satellite~\citep{PCP2018}. We will henceforth refer to this set of parameters as the {\it Planck} cosmology.

Observations of the `late' Universe (which for CMB scientists includes everything that happens well after recombination, $z \ll 1000$), generally agree with the {\it Planck} cosmology. However, differences between observations of the 'early' and 'late' Universe have generated significant interest, as they may suggest the presence of physics beyond the $\Lambda {\rm CDM}$ model. Local measurements of the expansion rate using the cosmic distance ladder~\citep{Freedman:2020dne,Riess:2021jrx} measure a higher value than predicted by the {\it Planck} cosmology at up to the $5 \sigma$ significance level. Independently, some local measurements of the matter distribution through weak gravitational lensing are also discrepant with \planck results~\citep{Amon:2022ycy}, although at a lower level of statistical significance.

However, it is a mistake to consider the ${\it Planck}$ parameter constraints a result of `early' Universe physics alone. Multiple effects affect the ${\it Planck}$ CMB power spectra at low redshift, which are commonly ignored in `early versus late' discussions. Of these effects, the most important are the Integrated Sachs-Wolfe (ISW) effect~\citep{Sachs1967}, Reionization, CMB Lensing; and Foregrounds, including galactic dust, point sources, cosmic infrared background (CIB), and the
Sunyaev-Zel’dovich (SZ) effect~\citep{Zeldovich:1969, Sunyaev:1972, Sunyaev:1980}.

% \begin{itemize}
%  \item Integrated Sachs-Wolfe (ISW) effect~\citep{Sachs1967}: The change in energy from photons propagated through an evolving gravitational potential. This happens during the start of matter domination (early ISW) and entering and during dark energy domination (late ISW). Of these, only the latter is considered a late Universe effect.
%  \item Reionization: free electrons after reionization scatter a fraction of the CMB photons by Thomson scattering. This effect is usually parameterized by a single parameter $\tau_{\rm re}$, measuring the optical depth to reionization.
%  \item CMB lensing: lensing alters the path of photons, remapping the last-scattering surface into the observed anisotropies.
%  Averaged over the whole sky, this smooths out the acoustic peaks of the CMB spectra, but the local variations produce a significant connected four-point function that can be used for lensing reconstruction. Since the smoothing effect is just the sky average of the local effect, a measured lensing reconstruction can be used to predict the amount of smoothing.

%  \item Foregrounds, including galactic dust, point sources, cosmic infrared background (CIB), and the
%  Sunyaev-Zel’dovich (SZ) effect~\citep{Zeldovich:1969, Sunyaev:1972, Sunyaev:1980}. These can be distinguished by frequency dependence, apart from the kinetic SZ (due to bulk peculiar motions of ionized gas) which is only a small signal for {\it Planck} and expected to have a very smooth spectrum.

% \end{itemize}

The late ISW signal is model-dependent, but is generically restricted to large scales and carries little statistical power due to cosmic variance. The lensing and foregrounds can be directly constrained empirically almost independent of cosmology: lensing reconstruction can be used to measure the lensing amplitude and hence the amount of power spectrum smoothing expected, and foregrounds (except kinetic SZ) can be subtracted or constrained using the non-blackbody scaling over multiple observed frequencies. By using these data constraints, and excluding the small sensitivity to the late ISW, it is therefore possible to model the observed CMB power spectra essentially independently of late-time physics. The late-time background evolution only affects the spectra via the angular diameter distance to last scattering. The angular size today of the comoving sound horizon at recombination is very accurately measured, and so encapsulates the only constraint on the late-time background evolution.

In this paper, we show how to build a CMB likelihood that does not require a cosmological model to model the late-time effects. Using this, we can robustly constrain cosmological models at early times independently of late-time physics, along with the comoving angular diameter distance that determines the observed angular size of the anisotropies.

Modelling the foregrounds empirically is standard practice, but the lensing is usually constrained using a full model.
Modelling the lensing independently of late-time structure growth is of particular interest due to the tensions with other data, suggesting possible inconsistencies in the $\Lambda$CDM model. There is also the possible $(2$--$3)\sigma$ preference of the \planck\ temperature data for more lensing-like smoothing than predicted by the \planck\ cosmology fit (high $\Alens$, see e.g. Refs.~\cite{Planck:2016tof,PCP2018,Efstathiou:2019mdh} for discussion). A large actual lensing signal is not supported by either the lensing reconstruction or the \planck\ polarization spectra, but if the apparent preference is not just a statistical fluctuation, it could indicate some modification to the early-Universe temperature CMB spectrum that could be explained by new physics.

Sometimes the lensing-scaling $\Alens$ parameter is varied as a free parameter, but, if a lensing likelihood is not also used, it allows lensing amplitudes that are ruled out by the lensing reconstruction. If lensing reconstruction is used, in general, it does not capture the full scale-dependence of possible variations in the lensing spectrum and amount of lensing smoothing. By using the lensing reconstruction data to fit the lensing smoothing empirically, we can constrain parameters without having to predict the cosmological dependence of the lensing signal at all.
An alternative would be to use the reconstructed lensing potential to delens sky at the map level~\citep{Green:2016cjr,Sehgal:2016eag,Yu:2017djs,Carron:2017vfg}, which should ultimately be more optimal than modelling the lensing at the power spectrum level. However, with current data this cannot be done perfectly, leaving some cosmology-dependent residual, and the delensed power spectrum can have complicated statistics and Gaussian biases that would need to be modelled~\citep{Carron:2017vfg}. Here we adopt the much simpler approach of just modelling the lensing power, which, as we shall show, is sufficient to provide good robust constraints, and can serve as a baseline for any delensing analysis. For Planck data, marginalizing over a free lensing amplitude $\Alens$ assuming late-time $\Lambda$CDM evolution gives similar results to allowing greater freedom in the lensing spectra. However, with future data, the CMB spectra are more sensitive to the relative amounts of lensing power on small scales, and marginalizing over a single parameter is then only sufficient to obtain robust constraints if the data are consistent with the $\Lambda$CDM prediction for the lensing spectral shape.

Previous work has tried to separate constraints of early and late cosmology and derive robust late-cosmology-independent parameter summaries ~\citep{Vonlanthen:2010cd,Audren:2012wb,Audren:2013nwa,Wang:2015tua,Motloch:2018pjy}.
Our approach is similar to ~\cite{Audren:2013nwa}, but we empirically constrain the lensing from the four-point information (as Refs.~\cite{Motloch:2018pjy,Motloch:2019gux}) rather than marginalizing out using only power spectrum information. We are also able to use the latest \planck PR4 lensing and power spectrum data to provide much more powerful constraints.

Future CMB data should be able to make a precision test of any new early-universe physics that could explain the Hubble tension. However, models that can do this typically also change the predictions for the late-time matter power, often in ways that appear inconsistent with the data. It's therefore possible that new physics is needed in both the early and late universe. By constraining the early-universe model in a way that's independent of the late-time structure, robust constraints can be placed on the early-universe model without making assumptions about the late-time evolution. Any detection of a violation of early-$\Lambda$CDM that is robust to the late-time evolution would be a powerful way to rule out the $\Lambda$CDM standard model using just CMB data, without making strong assumptions about the late-universe cosmology or the complexities of astrophysical observables. Parameter constraints from early-$\Lambda$CDM would also provide parameter consistency bounds for any model of the late cosmology that leaves the early-universe physics unmodified.

We start in Sec.~\ref{sec:methods} by describing the various late-time effects that can modify the observed CMB, and how we model, neglect, or constrain them. Then Sec.~\ref{sec:results} gives early-universe constraints on parameters from the latest Planck data, and an example simple demonstration for possible future data.

\section{Methods}\label{sec:methods}

\subsection{Integrated Sachs-Wolfe Effect}

\begin{figure}[t]
 \begin{center}
 \includegraphics[width=\columnwidth]{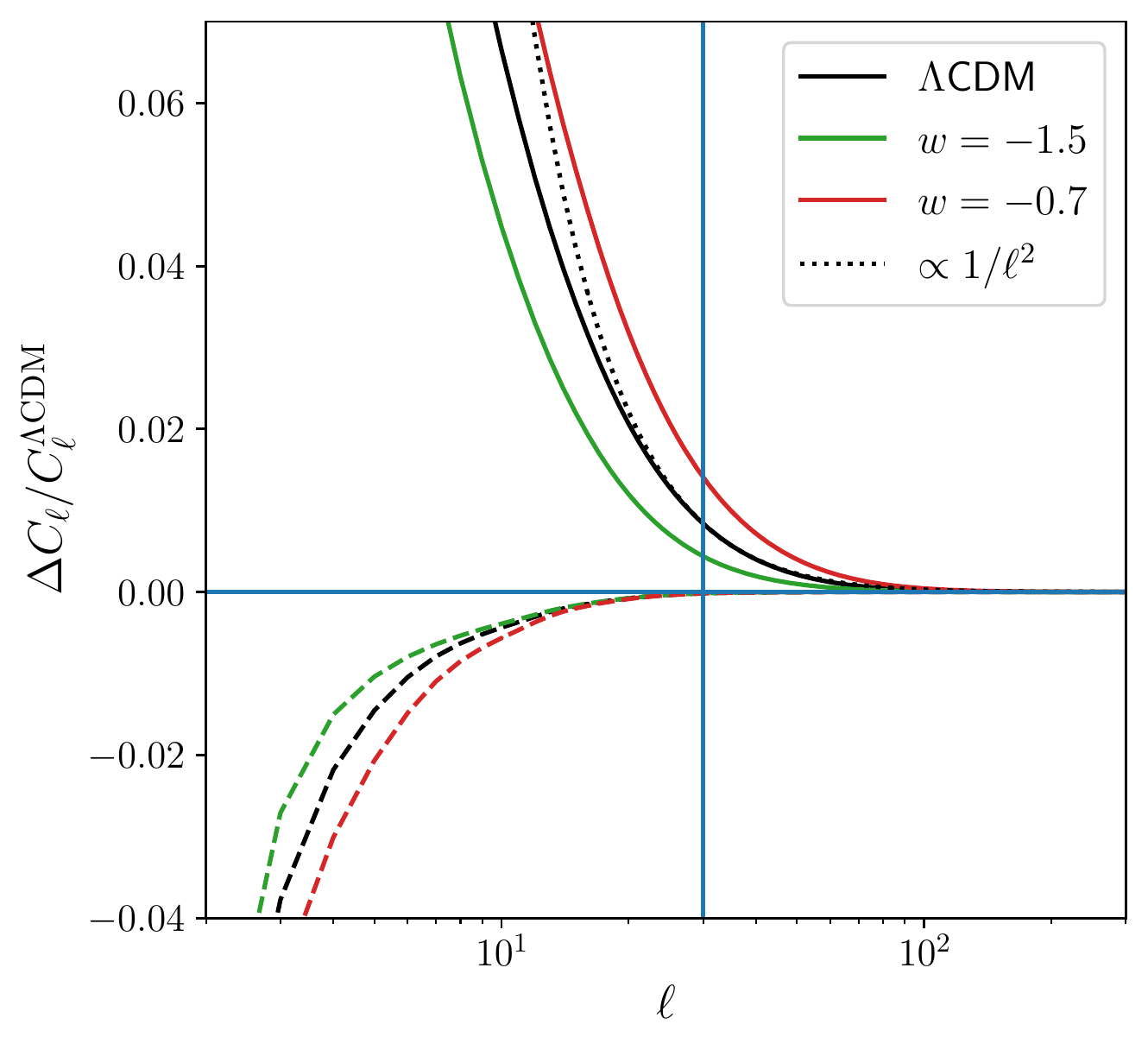}
 \caption{
 The contribution of the late ISW auto spectrum (solid) and cross-correlation of late ISW with the primordial signal (dashed), as a fraction of the $\Lambda$CDM temperature power spectrum for models with a different constant equation of state (at fixed early-universe parameters and acoustic angular scale $\theta_*$). The vertical line shows the lower multipole included in the high-$\ell$ \planck\ likelihood, and the dashed line a simple $1/\ell^2$ fit to the shape of the small residual ISW auto spectrum above this cutoff.
 Here ``late ISW'' is the contribution to the CMB anisotropy from redshift $z < 30$ in example fluid dark energy models with constant $w\equiv P/\rho$.
 }
 \label{fig:ISW}
 \end{center}
\end{figure}

The ISW consists of the change in energy from photons propagated through an evolving gravitational potential. This happens during the start of matter domination (early ISW) and entering and during dark energy domination (late ISW). Of these, only the latter is considered a late Universe effect.
%As discussed in the introduction, there are two types of ISW, but only one is affected by late-time physics.
This late-time ISW signal is imprinted on the scale of the evolving gravitational potentials that generate it. On small scales, there is a near cancellation between multiple perturbations along the line of sight, so the signal is only important on relatively large scales. On the scale of the current horizon there is some correlation between the ISW and the primordial perturbations, but at $\ell \gtrsim 30$ the perturbations are spatially separated and hence the ISW signal is essentially uncorrelated, contributing to the temperature power spectrum additively (see Fig.~\ref{fig:ISW}).
We can conservatively cut the temperature spectrum at $\ell < 30$ to remove the main signal. On smaller scales, we use an additive template with an amplitude decaying with $\ell(\ell+1)C_\ell \propto 1/\ell^2$, which, as shown in Fig.~\ref{fig:ISW}, is a reasonable fit to the rapidly-decaying signal in toy models with different constant dark energy equation of state parameter values.
For any plausible extended model with an ISW signal allowed by the very low-$\ell$ data, the residual at $\ell\ge 30$ must be small and hence does not need to be modelled accurately given the substantial cosmic variance in this region. At multipoles $\ell\ge 30$, we therefore model the total spectrum as the sum of the ISW template with a free amplitude, plus the theoretical prediction with ISW set to zero for redshifts $z<30$.

\subsection{Reionization}

%\begin{figure}[t]
% \begin{center}
% \includegraphics{tau}
% \caption{The effect of varying the optical depth to reionization in the CMB power spectrum.
% \AL{do we need it?}
% }
% \label{fig:tau}
% \end{center}
%\end{figure}

The effect of reionization (free electrons after reionization scatter a fraction of the CMB photons by Thomson scattering) on the high-$\ell$ CMB power spectra can be described by a single parameter, the optical depth to reionization $\tau_{\rm re}$. This quantifies how the amplitude of the high-$\ell$ CMB spectrum is suppressed, by a factor $e ^ {-2 \tau_{\rm re}}$. If the primordial power spectrum amplitude is $A_{\rm s}$, the high-$\ell$ CMB then only constrains the parameter combination $A_{\rm s}e^{-2\tau_{\rm re}}$.

On scales comparable to the horizon size at reionization, a large-scale polarization signal is generated (and a corresponding less-important contribution to the temperature). This allows $\tau$ to be constrained directly from the large-scale \planck polarization data. If the reionization history is known as a function of redshift, this signal does depend on the late-time cosmology. However, the reionization history is not known in any detail, and at \planck\ sensitivity there is very little time-evolution information available in the polarization power spectrum beyond the amplitude determined by $\tau$. Keeping $\tau$ as a free parameter can therefore model this signal with little cosmology dependence.

Reionization happens at redshift $z > 6.5$~\cite{Fan:2006dp}, and the contributions to the optical depth decline with scale factor as the free electrons dilute with the expansion of the Universe. Late-time dark energy effects on perturbation growth are therefore expected to have minimal impact. The effect of reionization on the temperature spectrum converges rapidly to $e^{-2\tau_{\rm re}}$ at high $\ell$, with deviation less than $1\%$ at $\ell \ge 30$ in $\Lambda$CDM. We therefore also neglect the small cosmology dependence of the effect of reionization on the temperature spectrum at $\ell \ge 30$. Any change in the ionization history at $z\gg 6$ should be included in the early-universe model.

These approximations should be very accurate for models where the background evolution is close to $\Lambda$CDM and modifications to perturbation growth are only important at $z\alt 6$.
With them, we can constrain $\tau$ (and hence $A_s$) using the polarization data, but we also quote final results for $A_s e^{-2\tau_{\rm re}}$ which is robustly measured by the high-$\ell$ spectra independently of the reionization history.

%Therefore, we can build marginalize over this effect, by simply marginalizing over all possible values of this parameter, as it is done in the {\it Planck} analysis. However, in the {\it Planck} analysis, we can constrain this parameter through low multipole information, especially in polarization, which breaks the degeneracy between $\tau{\rm re}$ and the amplitude of the primordial power spectrum. In this work, given that we are not using CMB low multipoles, we will get no constraints on $\tau_{\rm re}$, and as a consequence we will also barely constrain the power spectrum amplitude.

\subsection{CMB Lensing}

CMB lensing is the most complex CMB late-time anisotropy to correct. Lensing alters the path of photons, remapping the last-scattering surface into the observed anisotropies.
Averaged over the whole sky, this smooths out the acoustic peaks of the CMB spectra, but the local variations produce a significant connected four-point function that can be used for lensing reconstruction. Since the smoothing effect is just the sky average of the local effect, a measured lensing reconstruction can be used to predict the amount of smoothing. It affects the acoustic peaks by smoothing them and transferring power to small scales, which we cannot cut if we want to maintain constraining power. The impact on cosmological parameters is large: neglecting lensing would bias parameters by many sigma at Planck sensitivity~\cite{Lewis:2005tp}. Freely modelling the lensing smoothing, or allowing for a single-parameter lensing amplitude freedom $\Alens$~\cite{Calabrese:2008rt}, can remove much of the model dependence at the expense of signal, but can bias other marginalized parameter posteriors due to parameter degeneracies, and increases parameter error bars more than necessary.

To separate the effect of CMB lensing, we use the fact that the CMB lensing power spectrum can be constrained in a nearly model-independent way using the power spectrum of quadratic estimators (or more optimal estimators) for the lensing potential~\citep{Okamoto:2003zw,PL2018}.
Since the CMB lensing redshift kernel is broad, we assume the spectrum can be modelled by a smooth function. We use a log-amplitude, log-multipole spline with five or six nodes depending on the multipole range of the data. The spline nodes are chosen to produce a good fit to the lensing spectrum in $\Lambda$CDM, but allow considerable scope for scale-dependent smooth variation. The spline multipole nodes are taken to be at $L = \{ 7, 44, 125, 600, 1600, 3100\}$, where the last bin is dropped if there is no constraint at high multipoles (as for Planck).
We adopt broad flat priors on the spline node amplitudes, with log amplitudes for $[L(L+1)]^2C_L^{\phi\phi}/2\pi$ in the interval $(-22, -14)$.
The spline model is both constrained by the lensing reconstruction data and used for a consistent lensed power spectrum prediction.

\subsection{Foregrounds}

Galactic foregrounds such as dust can largely be separated from the CMB at the map level by using their frequency dependence.
Although extragalactic foregrounds are in principle predictable given a cosmological model, for current analyses they are usually either foreground cleaned or modelled empirically via smooth spectral templates. Neither of these approaches introduce late-time cosmological dependence, and the consistency of the empirical foreground templates can easily be checked via frequency differences.
Foreground cleaning on noisy maps will leave some residual foreground signal, but this can also be modelled empirically using smooth spectral templates.

\section{Results}\label{sec:results}

\begin{figure}[t]
 \begin{center}
 \includegraphics[width=.99\columnwidth]{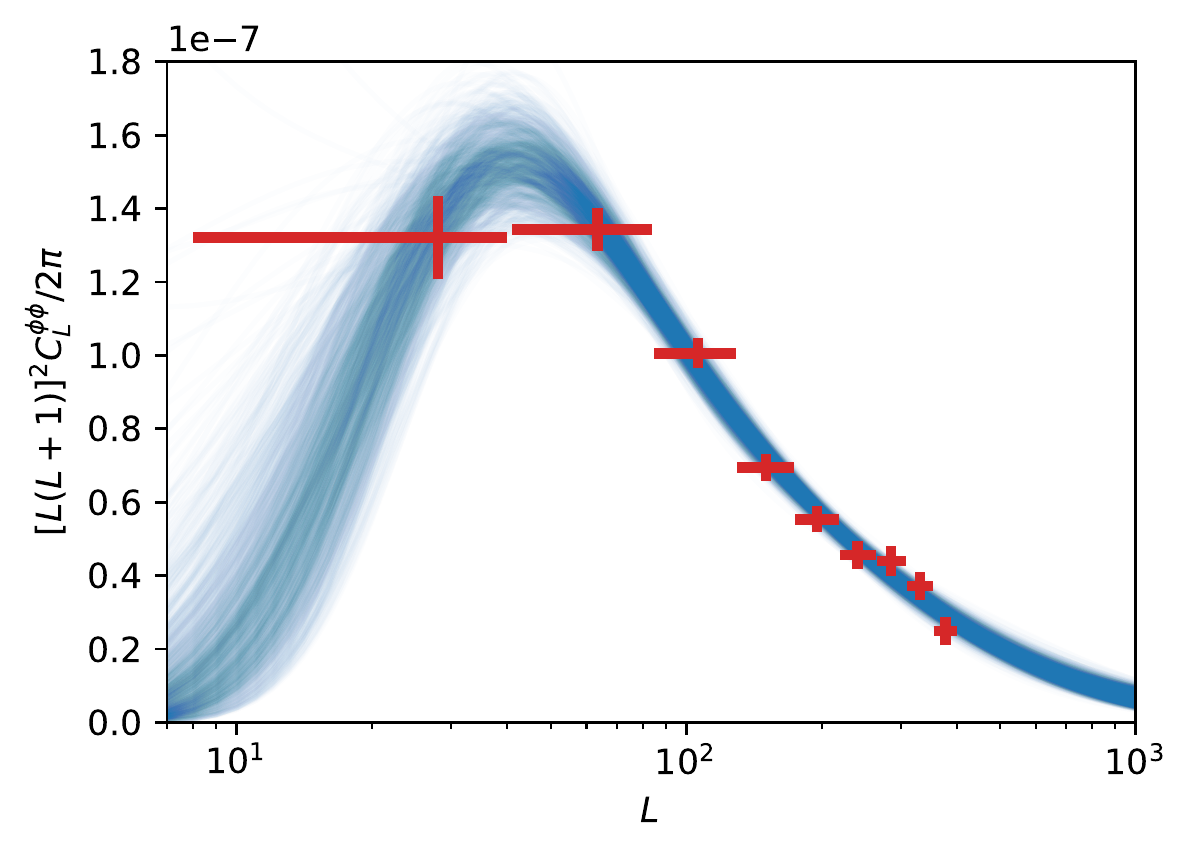}
 \caption{Planck constraints on smooth log spline fits to the CMB lensing power spectrum. The lines show 1000 samples from the posterior when varying five spline node amplitudes and early-universe $\Lambda$CDM parameters. Errors bars show multipole range and $\pm 1\sigma$ error for the conservative bandpowers from the Planck PR4 lensing analysis~\cite{Carron:2022eyg}.
 }
 \label{fig:template_results}
 \end{center}
\end{figure}

We use the {\it Planck} PR4 {\tt CamSpec} likelihood~\citep{Rosenberg:2022sdy}, implemented for {\tt cobaya}~\citep{Torrado:2020dgo}.
The {\tt Camspec} likelihood~\citep{Efstathiou:2019mdh,Rosenberg:2022sdy} uses the new PR4 (NPIPE) \planck maps~\citep{Akrami:2020bpw}, which use $\sim 8\%$ more data than the PR3 (2018) release, and a larger sky area than the 2018 baseline likelihood. As such we obtain tighter constraints than previous results for late-time independent parameter constraints, even without the improvements in lensing modelling.

The PR4 likelihood uses 143 and 217 GHz maps cleaned of galactic dust using the high-frequency \planck channels where the dust is much brighter. Residual foreground power is fit with a set of power law templates with free amplitudes and exponents for each frequency combination.
The likelihood only uses CMB power spectra at $\ell \ge 30$, and we do not use the low-$l$ TT likelihood to avoid sensitivity to very low-$l$ ISW.
We do use the low-$\ell$ EE likelihood~\citep{Planck:2019nip} to constrain the reionization optical depth.
The \planck PR4 lensing-only likelihood~\citep{Carron:2022eyg} is used to constrain the lensing template shape almost independently of cosmology and is constructed on foreground-cleaned maps.
Although the lensing reconstruction relies on fiducial CMB power spectra, it can be corrected self-consistently for different spectra~\cite{Ade:2015zua} and does not depend on the model used to generate the spectra. In practice, the observed CMB spectral shape is very well constrained by the data empirically, so there is almost no model dependence from the CMB power spectra when the lensing likelihood is combined with CMB power spectrum measurements.

We use {\tt Cobaya}\footnote{
\url{https://cobaya.readthedocs.io/en/latest/}} code~\citep{Torrado:2020dgo} to MCMC sample~\cite{Lewis:2013hha} cosmological parameters in a $\Lambda$CDM cosmology using theoretical predictions from {\tt CAMB}\footnote{\url{https://camb.info}}~\cite{Lewis:1999bs}, and analyse the samples with {\tt GetDist}\footnote{\url{https://getdist.readthedocs.io/}}~\cite{Lewis:2019xzd}. As an example, we run parameter chains in a $\Lambda$CDM cosmology, but by only reporting the late-cosmology-independent parameters, we can extract results that only depend on $\Lambda$CDM being valid at early times (as Refs.~\cite{Audren:2012wb,Audren:2013nwa}).
In extended models with new early-universe physics, results would have to be re-run to capture the changed early evolution, but the early-universe parameters would still be robustly independent of the late-time cosmology.

For each point in parameter space sampled by the chain, the lensing power spectrum spline amplitudes are used to construct the lensing power spectrum. This is then used to lens the model's prediction for the unlensed CMB power spectra, and the calculated lensed spectra are then used as the theory model for the CMB likelihoods. The lensing likelihood is used to constrain the spline model for the lensing spectrum and is corrected at leading order for sensitivity to CMB power by the lensed CMB power spectra.
Samples from the distribution of lensing power spectrum shapes are shown for Planck in Fig.~\ref{fig:template_results}. As expected, we obtain lensing spectra that are consistent with the data constraint, with smoothness ensuring broadly plausible shapes at all multipoles where it is not well constrained by the data. For Planck, the lensing smoothing effect is mostly determined by the lensing power up to intermediate scales $\ell \sim 200$, with little sensitivity to low or very high multipoles, and hence is well constrained empirically by the lensing data. For future data, there is more sensitivity to small-scale power in the lensed CMB damping tail, but the lensing reconstruction constraint will also be much tighter there.

\subsection{Planck early universe constraints}

\begin{figure}[t]
 \begin{center}
 \includegraphics[width=.99\columnwidth]{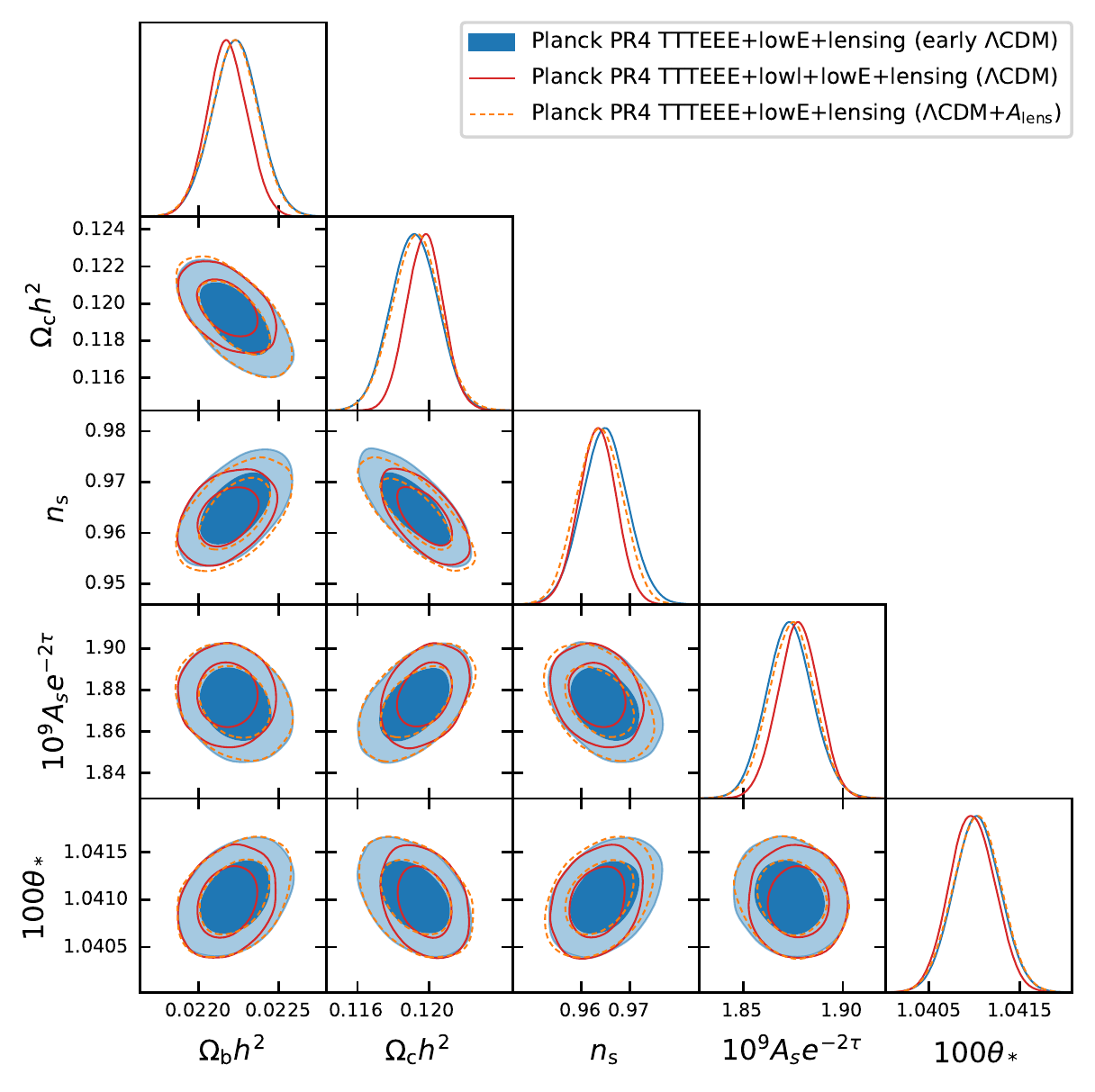}
 \caption{Marginalized constraints on early-$\Lambda$CDM cosmological parameters using the {\it Planck} PR4 likelihoods (filled contours), compared to the baseline result
 using full $\Lambda$CDM modelling (red contours). Contours contain 68\% and 95\% of the posterior probability, and $\theta_*$ is the ratio of the sound horizon to the angular diameter distance, which is calculated in the chain using background $\Lambda$CDM cosmology, but by construction is independent of the late-time model when measured using fixed CMB data.
 The dotted curves show the result using $\Lambda$CDM+$\Alens$, which for the Planck case are very similar to the early $\Lambda$CDM results.
 }
 \label{fig:posterior}
 \end{center}
\end{figure}

\begin{table}[]\centering
\begin{tabular} { l | c| c}
 Parameter & Early-$\Lambda$CDM & $\Lambda$CDM \\
\hline
{$\Omega_\mathrm{b} h^2$} & $0.02223\pm 0.00015 $ & $0.02218\pm 0.00013 $ \\
{$\Omega_\mathrm{c} h^2$} & $0.1192\pm 0.0013 $& $0.1198\pm 0.0012 $\\
$10^9 A_s e^{-2\tau}$ & $ 1.873\pm 0.012 $& $1.877 \pm 0.010$ \\
{$n_\mathrm{s} $} & $0.9648\pm 0.0047 $& $0.9633\pm 0.0039 $ \\
$100\theta_* $ & $1.04103\pm 0.00026 $& $1.04097\pm 0.00025 $ \\
\hline
$H_0 [{\rm km} s^{-1} {\rm Mpc}^{-1}] $ & $67.49\pm 0.58 $& $67.22\pm 0.45 $\\
\hline
\end{tabular}
\caption{Planck PR4 TTTEEE+lowE+lensing early-$\Lambda$CDM constraints on cosmological parameters, which are robust to changes in late-time structure growth and background evolution. The $\Lambda$CDM constraints use the full Planck PR4 TTTEEE+lowE+lowl+lensing data combination.
 The lower section Hubble parameter constraint is additionally assuming $\Lambda$CDM background evolution (but not growth of structure) to relate $\theta_*$ (the ratio of the sound horizon to the angular diameter distance) to the expansion rate today. The $\Lambda$CDM model approximates the neutrino masses as a single eigenstate of mass $m_\nu = 0.06{\rm eV}$.
\label{table:planck_pars}
}
\end{table}

As a demonstration with current data,~\cref{fig:posterior} shows \planck\ constraints on the parameters of an early $\Lambda$CDM model, along with the $\theta_*$ parameter that determines the observed angular size of the sound horizon (which is also constrained well independently of the late-time evolution).
By construction, the early-$\Lambda$CDM constraints use less data, so the constraints are slightly weakened compared to modelling the full evolution and lensing assuming $\Lambda$CDM at all times. However, the increase in error bars is modest, and these parameters are still very tightly constrained. The robust numbers given in Table.~\ref{table:planck_pars} could be used as early-universe model priors when constructing any variations in the late-time cosmology. In extended early-universe cosmologies, parameters that quantify the deviation from $\Lambda$CDM could also be constrained, with a detection away from the $\Lambda$CDM value confirming new physics independently of the evolution at late times.

To check the robustness of the early-$\Lambda$CDM constraints we also run a chain with a varying constant dark energy equation of state parameter $w$, which does change the late-time evolution but leaves high-redshift early-$\Lambda$CDM evolution unmodified. As expected, the results agree with those in Table.~\ref{table:planck_pars} up to shifts of the size of the small Monte Carlo error. In all cases, the ISW template contribution is empirically constrained to be $\alt 0.16$ ($2\sigma$) of the primordial signal at $\ell=30$, and has very little impact on results.

\subsection{Planck constraints on background $\Lambda$CDM}

The late-time evolution is now quite tightly constrained at the background level by supernovae, baryon acoustic oscillations and more local measurements.
Some classes of extended dark energy models can match $\Lambda$CDM background evolution accurately, only modifying the growth of structure.
In this case, we can use the same method as before, but can now also convert $\theta_*$ into $H_0$ (or equivalently $\Omega_m$), constraining it from the CMB independently of the late-time structure growth. With the same \planck\ data as the other parameters in Table.~\ref{table:planck_pars} we obtain $H_0 = (67.49 \pm 0.58) \ {\rm km} s^{-1} {\rm Mpc}^{-1}$. This is slightly higher, and with a slightly larger error, than the result obtained assuming $\Lambda$CDM structure growth at low redshift. However, the shift is small, suggesting uncertainties in the physical lensing amplitude modelling cannot explain $\Lambda$CDM discrepancies with the local distance ladder measurement result $H_0 = (73.15 \pm 0.97) \ {\rm km} s^{-1} {\rm Mpc}^{-1}$ ~\citep{Riess:2022mme}, which remains discrepant at $\sim 5.0 \sigma$ (down from $5.5\sigma$ assuming $\Lambda$CDM structure growth). The weaker tension with other data (e.g. Ref.~\cite{Freedman:2020dne}) would be similarly little changed.

Uncertainties in late-time evolution also arise because of non-linear modelling even in a full $\Lambda$CDM model. However, for CMB lensing, which is a high-redshift probe where the perturbations are mostly fairly linear, the non-linear modelling uncertainty is restricted to fairly small scales in the lensing spectrum. As such, the sensitivity of the CMB spectra to non-linear evolution is modest. For the lensing spectrum, the impact is more important but can be easily modelled robustly using more physical parameterizations of the effect (e.g.~\cite{Baldauf:2016sjb}), rather than the very conservative approach adopted here which assumes no theoretical knowledge of the late-time evolution.

%
%\begin{figure}[t]
% \begin{center}
% \includegraphics[width=.99\columnwidth]{posterior_w.pdf}
% \caption{Posterior distributions on the dark energy equation of state, the cold dark matter density, and the Hubble parameter (in units of ${\rm km/s/Mpc}$) using our early Universe likelihood.
% \AL{I would say we are more interested in showing that the late-indep parameters are unchanged [robust to late-time DE model], rather than showing the actual constraints on H0/w (which are entirely determined by the priors).}
% }
% \label{fig:posterior_w}
% \end{center}
%\end{figure}

%As previously described, we use a varying dark energy equation of state to verify that our Early Universe only likelihood does indeed use early-time physics only. As shown in~\cref{fig:posterior_w}, our likelihood cannot constrain the dark energy equation of state, except for very high values, which would start to have an impact in the early Universe. Furthermore, we see that varying $w$ does not affect constraints on other parameters (the figure shows this only for the cold dark matter density, but the same is true for all other parameters shown in~\cref{fig:posterior}) except for the Hubble constant $H_0$. This is because, as previously discussed, changing the dark energy equation of state does change the comoving angular diameter distance, and therefore the Hubble parameter. Therefore, this figure does show that our likelihood is independent of late-time physics.

\subsection{Future CMB constraints}

\begin{figure}[t]
 \begin{center}
 \includegraphics[width=.99\columnwidth]{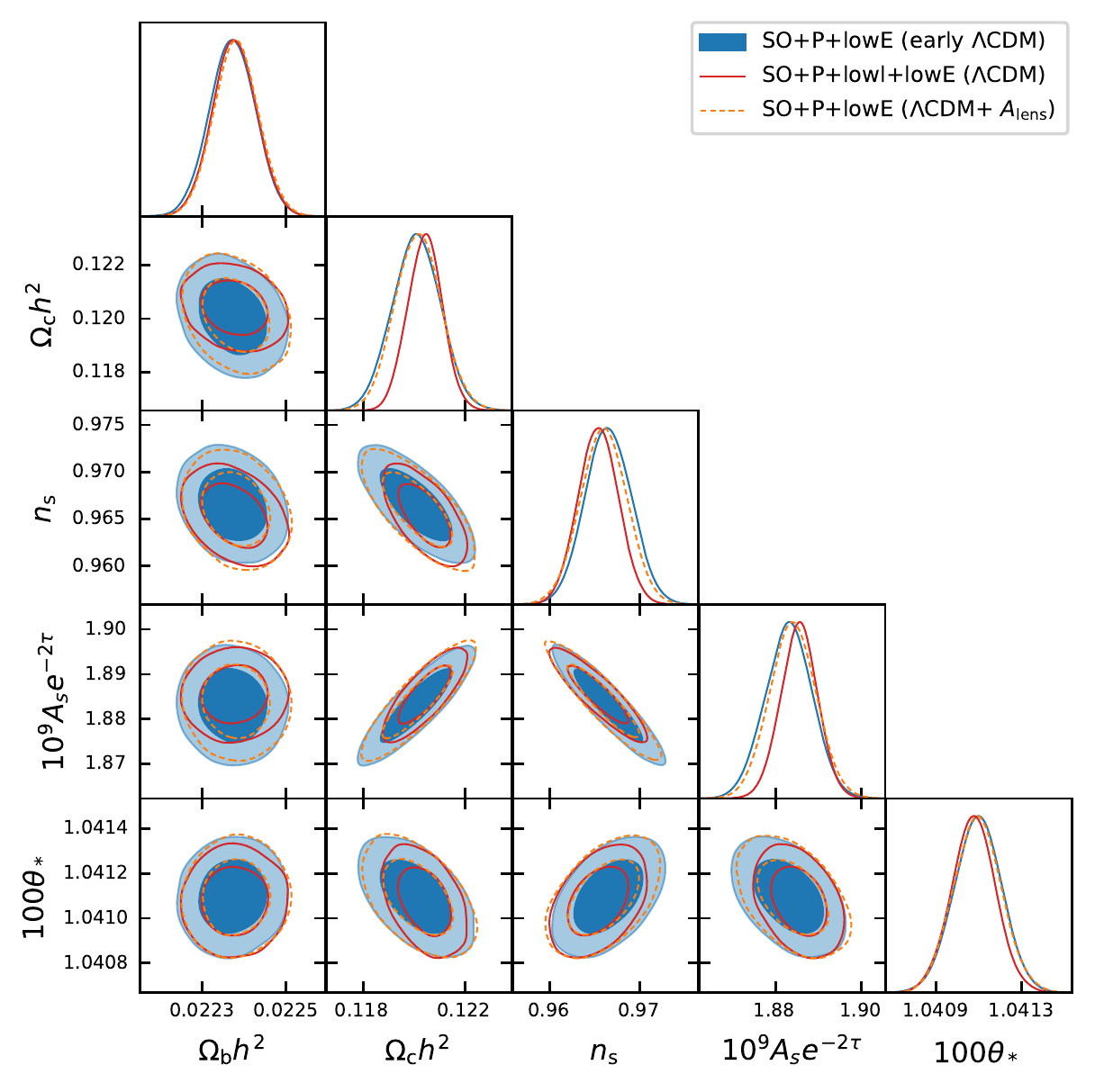}
 \caption{Forecast marginalized constraints on early-$\Lambda$CDM cosmological parameters using a simple SO+Planck model (filled contours), compared to the baseline result
 using full $\Lambda$CDM modelling (red contours).
 Here ``SO+P" refers to fiducial Simons Observatory data over $40\%$ of the sky plus fiducial Planck data over 30\% of the sky at $\ell\ge 30$,
 ``lowl" is a fiducial Planck likelihood at $\ell < 30$ over $80\%$ of the sky,
 and all results include the full Planck low-$\ell$ EE constraint.
 The dashed lines show the constraints assuming full $\Lambda$CDM evolution but with a single free lensing amplitude parameter $\Alens$, which are similar
 in this case, since we are assuming fiducial data that match the shape of the $\Lambda$CDM lensing spectrum prediction.
 The inclusion or otherwise of ``lowl'' does not have a large effect, since with SO data constraints on power law power spectra are dominated by smaller-scale information.
 }
 \label{fig:SOposterior}
 \end{center}
\end{figure}

The Planck CMB power spectrum data, which do not include the B-mode power, are only sensitive to one lensing amplitude parameter~\cite{Smith:2006nk}. As such, very similar constraints can be obtained by dropping low-$\ell$ TT data and calculating the lensing spectrum in $\Lambda$CDM, and scaling the result by a free lensing amplitude parameter $\Alens$ relative to the $\Lambda$CDM prediction. The lensing reconstruction data constrain the lensing amplitude on a similar scale to that relevant to the CMB spectral smoothing, so $\Alens$ is empirically constrained. However, future data starts to constrain the CMB damping tail, where the impact of lensing becomes more sensitive to the small-scale spectral shape, and the lensing spectrum will have more constraining power on scales smaller than those relevant for the main smoothing effect. If B modes are included, they also have more sensitivity to smaller scales.

As a simple example, we consider temperature and E-mode polarization parameter constraints from Simons Observatory~\cite{Ade:2018sbj}, using fake data given by theoretical $\Lambda$CDM spectra and evaluating simple mean log-likelihoods scaled to the expected $\sim 40\%$ sky area. The B-mode lensing is expected to be entirely lensing, apart from a possible small tensor signal on large scales, so we do not consider it.
We take foreground-cleaned noise curves from the SO website\footnote{\url{https://github.com/simonsobs/so_noise_models}}.
It is unclear to what high multipole the temperature likelihood and foreground residual model can be relied on. Here we take $l_{\rm max}=4000$ with no residual modelling, which represents a worse case in terms of sensitivity to the high-$L$ lensing spectrum.

There is a covariance between the lensed CMB power and the lensing reconstruction power, which should be included in the likelihood model to avoid double counting lensing information~\cite{Schmittfull:2013uea,Peloton:2016kbw,Motloch:2016zsl}. We include the dominant terms from the cosmic variance of the lenses:
\begin{equation}
 \text{cov}\left( \hat{C}^{\phi\phi}_{L_1}, \hat{C}^{\tilde U\tilde V}_{l_2}\right)\approx f_{\rm sky}^{-1} \sum_{L_3} \frac{\partial \hat{C}^{\phi\phi}_{L_1}}{\partial C^{\phi\phi}_{L_3}} \frac{2}{2L_3+1}(C_{L_3}^{\phi\phi})^2 \frac{\partial C_{l_2}^{\tilde U\tilde V}}{\partial C^{\phi\phi}_{L_3}}.
\end{equation}
The lensing estimator $\hat{C}^{\phi\phi}_{L_1}$ responds to the lensing spectrum directly, and via an $N^{(1)}_L$ bias term. The $N^{(1)}_L$ term arises from non-primary contractions, and is linear but non-diagonal in the lensing spectrum~\cite{Kesden:2003cc}. The dependence of  $N^{(1)}_L$ on the lensing spectrum can therefore be encoded in a matrix giving the linear mapping, and we calculate this matrix the flat sky approximation using the {\sc lensitbiases} code\footnote{\url{https://github.com/NextGenCMB/lensitbiases}}~\cite{Carron:2022eyg}. This only contributes a small correction. The remaining derivatives of the lensed CMB spectra, $C_{l_2}^{\tilde U\tilde V}$, are calculated using
{\tt camb}\footnote{\href{https://camb.readthedocs.io/en/latest/correlations.html\#camb.correlations.lensed_cl_derivatives}{camb.correlations module}}.
The CMB lensing auto-covariance also has an off-diagonal contribution calculated using the $N^{(1)}_L$ coupling matrix.
Including these correlations does not have a large effect on the main $\Lambda$CDM parameters, but does significantly weaken constraints on the lensing spectrum, especially on small scales.
It is also important to include when running $\Lambda$CDM$+\Alens$ models, to avoid artificially tight constraints on the lensing amplitude.
The covariance does depend on a fiducial lensing model, however the fiducial model can be fit to the empirical lensing data, and any remaining model dependence of the covariance should be a small correction on a correction.

Forecast results are shown in Fig.~\ref{fig:SOposterior}. Although discarding the late-time information does increase error bars, there are still good robust constraints. Of the well-constrained parameters, the matter density parameter is most affected, because this parameter gains the most from having lensing information. Lensing information also helps to separate the primordial amplitude from the optical depth, so removing the lensing information increases the error bar on $A_s$ more significantly from $\pm 1.9\times 10^{-11}$ ($\Lambda$CDM) to $\pm3.4\times 10^{-11}$ (early $\Lambda$CDM).

As a test analysis of a non-standard model where the late-time physics varies, Fig.~\ref{fig:SOposterior} uses fiducial data generated in a model with late-time dark energy-dark matter coupling. This gives a significantly different lensing spectrum while leaving the early-universe evolution unchanged.
Specifically, we use the specific simple model of Ref.~\cite{Asghari:2019qld,Poulin:2022sgp} with $\Gamma_{\rm DMDE}/(H_0\rho_c)=2.5$, implemented in {\tt CAMB}.
As expected, our early-$\Lambda$CDM analysis recovers the same results independent of the late-time growth. However, if one simply models the data using $\Lambda$CDM+$\Alens$ the result is biased, for example recovering $n_{\rm s}$ with a bias of about $0.2 \sigma$, since $n_{\rm s}$ is sensitive to the amount of lensing power in the CMB damping tail. The varying $\Alens$ model is therefore not sufficient in cases where the shape of the lensing spectrum can differ significantly from $\Lambda$CDM.
The model of Ref.~\cite{Poulin:2022sgp} that we used is of course extreme, and would likely be clearly detected or ruled out by redshift distortion, lensing, and other data. The $\Lambda$CDM+$\Alens$ model may remain a useful proxy for cases where the measured lensing results are close enough to $\Lambda$CDM that the single lensing amplitude parameter encapsulates the posterior freedom sufficiently well.

\begin{figure}[t]
 \begin{center}
 \includegraphics[width=.99\columnwidth]{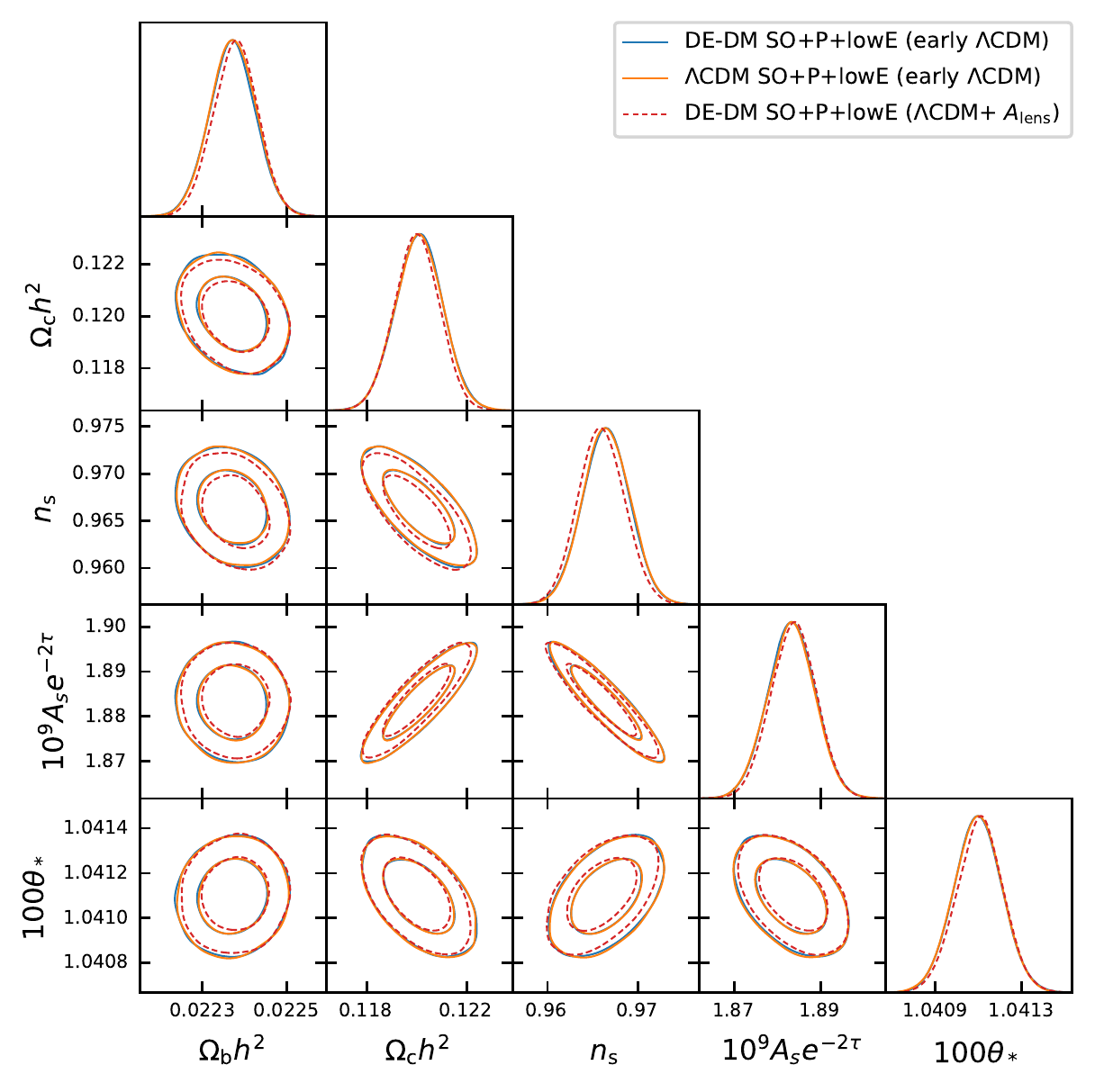}
 \caption{Forecast early-$\Lambda$CDM cosmological parameter constraints when the model data has a dark-energy/dark-matter (DE-DM) coupling at late times, otherwise using the same configurations as Fig.~\ref{fig:SOposterior}. The solid contours are virtually identical, so our early-$\Lambda$CDM analysis recovers consistent results independent of the late-time evolution as expected. Instead modelling the data using $\Lambda$CDM+$\Alens$ gives the slightly biased dashed contours: a single lensing amplitude parameter is not sufficient to robustly constrain the early-universe if the lensing spectral shape is significantly different from $\Lambda$CDM expectations.
 }
 \label{fig:SODEDM}
 \end{center}
\end{figure}

\section{Conclusions}

We updated previous work on measuring early-universe parameters independently of the late-time model, using the latest cosmic microwave background data, and making full use of the available empirical information about CMB lensing. With less information, constraints are weaker than when assuming a full model, but only to a modest extent since the effect of lensing is empirically constrained by lensing reconstruction. Our simple parameterized fitting method can serve as a baseline reference for future work developing more optimal map-level delensing approaches. Although we only demonstrated it explicitly on early-$\Lambda$CDM models, the same approach could be used with extended models (for example, models that modify the early-universe physics in an attempt to shift the sound horizon and hence explain the current tension in $H_0$ measurements).

With future data, the separation of early and late-time information can be a powerful way to test cosmological models of early evolution independently of late-time modelling assumptions, and is a conservative complement to full analyses assuming specific model evolution.

\section*{Acknowledgements}

We thank Erminia Calabrese for her suggestions.
We acknowledge support by the UK STFC grant ST/T000473/1.



\bibliography{example,texbase/antony,texbase/cosmomc} % if your bibtex file is called example.bib

%%%%%%%%%%%%%%%%%%%%%%%%%%%%%%%%%%%%%%%%%%%%%%%%%%

%%%%%%%%%%%%%%%%% APPENDICES %%%%%%%%%%%%%%%%%%%%%

\end{document}